\documentclass[twocolumn,shortnote]{jpsj2} 
\title{Berry's phase \\ in the multimode Peierls states}

\author{Tohru Kawarabayashi,
Yoshiyuki Ono
and Chiduru Watanabe
}

\inst{Department of Physics and Research Center for Materials with Integrated Properties, Toho University, 
Miyama 2-2-1, Funabashi, Chiba 
274-8510, Japan \\

}

\kword{Peierls states, two-dimensional systems, Berry's phase}

\begin{document}
\maketitle

In the present short note, we show that Berry's phase
associated with the adiabatic change of local variables 
in the Hamiltonian \cite{Berry,Hatsugai,MH} 
can be used to characterize  
the multimode Peierls state, which has been proposed as a new type 
of the ground state 
of the two-dimensional(2D) systems with the electron-lattice 
interaction.\cite{OH,HO} It is called ``multimode'' since  
its lattice distortion 
pattern consists of more than one Fourier components.
It has been shown \cite{OH,HO} that the state is  
the ground state of the half filled 
tight-binding model on the 2D square lattice described by the 
Hamiltonian
\begin{equation}
\begin{split}
 H  =  \sum_{\mib{r}} &  
 \{-(t_0 - \alpha v_x(\mib{r}))(c_{\mib{r}}^{\dagger}
 c_{\mib{r}+\mib{e}_x} + {\rm h.c.}) \\ 
  & - (t_0 - \alpha v_y(\mib{r}))(c_{\mib{r}}^{\dagger}
 c_{\mib{r}+\mib{e}_y} + {\rm h.c.})  \\
  & + \frac{K}{2}(v^2_x(\mib{r}) + v^2_y(\mib{r})) \}.
\end{split}
\label{hamiltonian}
\end{equation}
Here the lattice distortion at the site $\mib{r}$ is denoted by 
$\mib{v}(\mib{r})=(v_x(\mib{r}),v_y(\mib{r}))$ and is defined by 
the difference of the lattice displacements $\mib{u}(\mib{r})$ as 
\begin{equation}
 v_{x(y)}(\mib{r}) = u_{x(y)}(\mib{r}+\mib{e}_{x(y)}) - u_{x(y)}(\mib{r}),
\end{equation}
where $\mib{e}_{x(y)}$ denotes the unit vector in the $x(y)$ direction.
In the present note, 
the spin degrees of freedom are neglected for simplicity.
The third  term in the Hamiltonian 
describes the elastic energy of the lattice within the 
harmonic approximation. In this model, the relevant parameter for the 
electron-lattice coupling is $\lambda = \alpha^2/(t_0 K)$.
This 2D Peierls model is a simple generalization of the  
Su-Schrieffer-Heeger model\cite{SSH,TLM} originally introduced 
in one dimension for the analysis of solitons in polyacetylene. 
The model has been widely adopted for taking into account the effect of 
the electron-phonon coupling in 2D systems.\cite{TH,YK}
The conventional Peierls instability expected from the shape of 
the Fermi surface is the instability with the wave vector
$(\pi,\pi)$.
It has been shown \cite{OH,HO}, however,  
that the state having the lattice distortion
pattern with the wave vector $(\pi, \pi)$ is not the true 
ground state of the model at half filling, and 
that the true ground state exhibits a 
complex distortion pattern with more than one Fourier components
including $(\pi,\pi)$, which is called the multimode Peierls state. 
With the additional components, which turn out to be  
parallel to $(\pi,\pi)$, the energy gap opens in the whole 
range of the Fermi surface. Note that for the lattice distortion 
described only by the $(\pi,\pi)$ mode, there exist gapless points at 
the Fermi surface. More interestingly, there are many multimode patterns 
giving the same ground state energy.\cite{HO} This huge degeneracy is removed 
by introducing an anisotropy,\cite{OW} for instance, by replacing the 
hopping amplitude $t_x =(t_0 -\alpha v_x(\mib{r}))$ and $t_y=(t_0 - \alpha
v_y(\mib{r}))$ with $(1+\kappa /2)t_x$ and $(1-\kappa/2)t_y$, 
respectively.
The parameter $\kappa (0<\kappa<2)$ determines the strength of anisotropy.
It has been pointed out that the real space characterization of the 
distortion patterns is useful to understand the relations between those 
degenerated distortion patterns.\cite{Chiba} 

In spite of these interesting properties of the Peierls instability in 
two dimensions, 
the property of the electron wave function has not been discussed 
in the present model. The reason is that 
the electron density is uniform and is exactly 
equal to $\langle n \rangle =1/2$
due to the electron-hole symmetry, irrespective of the lattice distortion 
patterns. Namely, the electron density does not show any structure 
related to the Peierls instability.
In this short note, we analyze, instead of the electron density, 
the geometrical quantum phase factor, 
originally discussed by Berry,\cite{Berry} 
associated with a cyclic evolution of  
an external parameter. 
We find that the geometrical phase, called Berry's phase, reveals 
interesting topological structure of the electron wave function 
of the multimode Peierls states.

In recent years, Berry's phase associated with the adiabatic change 
of the local phases of the Hamiltonian 
has been proposed as a topological
order parameter for a quantum ground state and 
applications to electron systems as well as spin systems have been 
demonstrated.\cite{Hatsugai,MH} It has been shown generally\cite{Hatsugai}
that Berry's phase associated with a closed loop $C$ in the parameter space 
is quantized to be $0$ or $\pi$ if 
the system is invariant under an antiunitary operation and the excitation 
energy gap is finite along the loop $C$. 
For fermion systems, an example of 
the antiunitary symmetry is the particle-hole 
symmetry and the quantization of  
Berry's phase has been confirmed for the 
random hopping model at half filling.\cite{Hatsugai}

With the above progress in the topological characterization of the 
quantum state, we examine the correspondence of the topological 
property of the electron wave function and the lattice distortion 
patterns in the case of the Peierls state in two dimensions. 
We adopt the Hamiltonian (\ref{hamiltonian}) and investigate the 
relationship between the multimode distortion patterns and the 
values of Berry's phase.
Following Hatsugai\cite{Hatsugai}, we 
consider Berry's phase associated with the cyclic 
change of the phase $\theta_{x(y)}(\mib{r})$ of the local hopping amplitude 
$(t_0 -\alpha v_{x(y)}(\mib{r}))=|t_0 -\alpha v_{x(y)}(\mib{r})|\exp({\rm i}\theta_{x(y)}(\mib{r}))$. 
In order to evaluate Berry's phase numerically, 
the discrete phases
$\theta_n = 2n\pi/N$ with $n=1,2,\ldots, N$
are introduced. Then Berry's phase $\gamma$ associated with the 
cyclic change of $\theta_{x(y)}(\mib{r})$ 
from $0$ to $2\pi$
can be expressed as \cite{Hatsugai,MH,FHS}
\begin{equation}
 \gamma  =  \arg \prod_{n=1}^N \det A_n/|\det A_n| 
\label{berry_def} 
\end{equation} 
with $ (A_n)_{ij} = \langle \phi_i(\theta_{n+1}) | \phi_j(\theta_n) \rangle$, 
where the $i$th filled eigenstate of electrons for the case of  
$\theta_{x(y)}(\mib{r}) = \theta_n$ 
is denoted by $|\phi_i(\theta_n)\rangle$. Since we consider the 
half-filled case, the number of electrons $N_e$ is $L^2/2$ in the case of 
$L \times L$ system. The matrix $A_n$ is therefore a 
$N_e \times N_e$ matrix and its $ij$ element is denoted by $(A_n)_{ij}$.
In the presence of the multimode Peierls distortion, 
all the states with negative 
energy are occupied and the energy gap exists at the Fermi energy $E_F=0$.
Berry's phase is therefore quantized to be $0$ or $\pi$ in the 
multimode Peierls state. Since Berry's phase we consider here is associated 
with the cyclic evolution of the phase of the local bond, its  
value depends on the position of the bond. 
We can therefore assign the value of Berry's phase to 
each bond of the square lattice.
In actual calculations, we first determine the lattice distortion of the 
ground state of the Hamiltonian 
(\ref{hamiltonian}) self-consistently.\cite{OH} 
We then evaluate Berry's phase by changing 
the phase of the local bond variable keeping the distortion fixed.
In this way, we obtain information on the topological aspect of the 
ground state wave function of the system (\ref{hamiltonian}).

\begin{figure}
\begin{center}
\includegraphics[width=6cm]{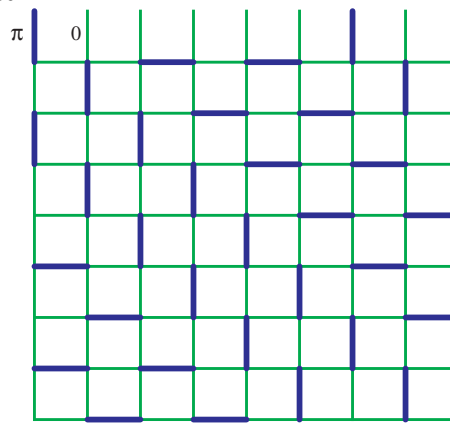}
\end{center}
\caption{(Color Online) Berry's phases in the multimode Peierls state.
The results for $L=8$ and $\lambda = 0.1$ with 
periodic boundary conditions are shown. 
The corresponding lattice distortions are shown in Fig. \ref{f2}.
Thick lines and thin lines 
represent the bonds with Berry's phase $\gamma = \pi$ and those 
with $\gamma =0$. The parameter $N$ is assumed to be $100$.} 
\label{f1}
\end{figure}

\begin{figure}
\begin{center}
\includegraphics[width=6cm]{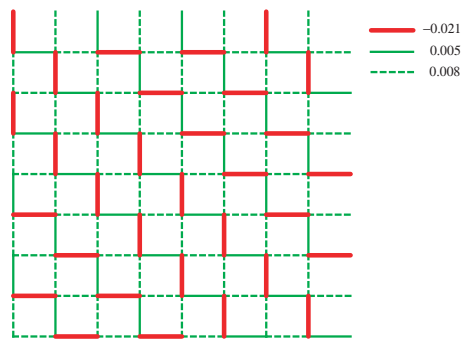}
\end{center}
\caption{(Color Online) 
An example of the lattice distortions of the multimode Peierls 
state with $(\pi,\pi)$ and $(\pi/4,\pi/4)$ modes. The parameters are  
$L=8$ and $\lambda=0.1$. Periodic boundary conditions are assumed. 
Thick solid lines represent the bonds with a negative lattice distortion.
Thin solid lines and dotted lines stand for positive lattice distortions.
The actual distortions for thick solid lines, thin solid 
lines and thin dotted lines 
are -0.021, 0.005 and 0.008 in units of $t_0/\alpha$, respectively.
The difference between the positive 
distortions for thin solid lines and 
for thin dotted 
lines vanishes in the thermodynamic limit ($L \rightarrow \infty$).}
\label{f2}
\end{figure}

Berry's phase evaluated for each bond 
and the Peierls distortion pattern are shown in 
Fig. \ref{f1} and Fig. \ref{f2}, respectively. 
Here it is clearly seen that Berry's phase is $\pi$ for every bond with
negative distortions and it is $0$ for every bond with 
positive distortions. This 
property is insensitive to the strength of the electron-lattice coupling 
$\lambda$. It is to be noted that in the case of $\lambda = 0.1$, the
changes in the hopping amplitudes due to the 
distortions are less than a few percent. 
Even for such a small distortion, quantized Berry's phases 
remain to be the same as those for larger distortions in the case of  
a larger coupling $\lambda$. This point is in contrast with the case of the 
random bond model \cite{Hatsugai} where the bond with 
a larger hopping amplitude does 
not necessarily coincide with the bond with $\gamma =\pi$.
We also perform the same analysis in the presence of anisotropy and
find that the bonds with $\gamma = \pi$ perfectly match to the bonds 
with negative distortion. It is expected therefore 
that only the sign of distortions
determines the values of quantized Berry's phase. 
The present correspondence  
suggests that a quantum transition must occur 
between different patterns of negative distortions.
The description by quantized Berry's phase is thus important 
for analyzing the transitions between the multimode states and the 
conventional Peierls state with the ($\pi,\pi$) mode in 
the presence of  
anisotropy.\cite{OW}

In conclusion, we have evaluated Berry's phases in the multimode 
Peierls states in two dimensions. We have shown that there exists one 
to one correspondence with the quantized value of Berry's phase 
and the sign of distortions.
This topological characterization by quantized Berry's phase reveals 
the topological stability of the multimode Peierls states, which was 
not obvious in the characterization by the lattice distortions, and  
is useful for the description of 
the multimode Peierls states with complicated lattice distortions.

\section*{Acknowledgment}
The authors thank Y. Hatsugai for useful discussions.

\appendix

\end{document}